\documentclass[%
reprint,
superscriptaddress,
showpacs,
preprintnumbers,
amsmath,
amssymb,
aps,
]{revtex4-1}
\usepackage[unicode=true,colorlinks=true,citecolor=blue,urlcolor=blue]{hyperref}

\usepackage{graphicx}
\usepackage{dcolumn}
\usepackage{bm}
\usepackage{colortbl}

\begin{document}
\newcommand{\addDima}[1]{\textcolor{red}{#1}}
\newcommand{\commentDima}[1]{\textcolor{red}{\textbf{Dima: }\textit{#1}}}

\title{Nonlinear light absorption in colloidal CdSe/CdS nanoplatelets}

\author{V.\,N.\,Mantsevich}\email{vmantsev@gmail.com}
\affiliation{Lomonosov Moscow State University, 119991 Moscow,
Russia}
\author{D.\,S.\,Smirnov} \affiliation{Ioffe Institute,
194021, St. Petersburg, Russia}
\author{A.\,M.\,Smirnov}
\affiliation{Kotel'nikov Institute of RAS, 125009 Moscow, Russia}
\author{A.\,D.\,Golinskaya}
\affiliation{Lomonosov Moscow State University, 119991 Moscow, Russia}
\author{M.\,V.\,Kozlova}
\affiliation{Lomonosov Moscow State University, 119991 Moscow, Russia}
\author{B.\,M.\,Saidjonov}
\affiliation{Lomonosov Moscow State University, 119991 Moscow, Russia}
\author{V.\,S.\,Dneprovskii}
\affiliation{Lomonosov Moscow State University, 119991 Moscow, Russia}
\author{R.\,B.\,Vasiliev}
\affiliation{Lomonosov Moscow State University, 119991 Moscow,
Russia}

\begin{abstract}
We investigated the nonlinear optical properties of CdSe/CdS
nanoplatelets in the vicinity of heavy hole and light hole exciton
resonances. The two color pump-probe technique was applied. The
first intense pulse created non-equilibrium exciton population,
which was detected as a decrease of probe light absorption. We
observed intense scattering of excitons between heavy- and
light-hole excitonic states. We also studied experimentally
saturation of absorption in nanoplatelets. Theoretical description
of these phenomena allowed us to determine parameters of exciton
dynamics in nanoplatelets.

\end{abstract}

 \keywords{} \maketitle

\section{Introduction}
Semiconductor nanocrystals attract a great deal of attention due to
their physical and chemical properties promising for optoelectronic
applications~\cite{Xu}. Nanocrystals can be conveniently grown by
colloidal synthesis, which allows to precisely control their
shape~\cite{Manna}, size~\cite{Murray} and crystal
structure~\cite{Michalet}. Recently, this approach was applied to
synthesize finite size semiconductor quantum wells, which are called
nanoplatelets (NPLs)~\cite{Ithurria}. NPLs demonstrate strong
quantum confinement of charge carriers, because their thickness
along $[001]$ axis is of the order of several
monolayers~\cite{Ithurria}, which can be used to tune the optical
absorption and photoluminescence
spectra~\cite{Hines,Talapin,Chen,Chen_1,Mahler}. NPLs exhibit a
large exciton binding energy~\cite{Grim,Achtstein} and demonstrate a
number of other remarkable properties such as narrow emission lines
even at room temperature, tunable emission wavelength, short
radiative lifetimes, giant oscillator strength, high quantum yield
and vanishing inhomogeneous
broadening~\cite{Ithurria_2,Mahler_1,Tessier,Biadala,Pelton}. This
makes NPLs very attractive for application in future optoelectronic
devices as bright and flexible light emitters~\cite{Chen,She}, as
colloidal lasers~\cite{Grim,Guzelturk} or for biomedical
labeling~\cite{Bruchez}.

Despite a set of potential applications of colloidal NPLs,
experimental investigation of their fundamental properties was
mainly focused on the time ranges of the excited states dynamics,
such as decay pathways of the single-exciton
state~\cite{Achtstein,Kunneman}, recombination dynamics of band edge
excitons~\cite{Biadala}, non-radiative Auger
recombination~\cite{Grim,Baghani} or photoluminescence decay
dynamics~\cite{Tessier_1,Olutas}. In the same time various possible
applications of NPLs in optical devices calls for detailed
investigation of their nonlinear optical properties at high (room)
temperature~\cite{Skott,Selyukov,Heckmann}. This however have hardly
been done up to date.

In this work, we study nonlinear transmission of CdSe/CdS NPLs with
different shell thicknesses. We observe bleaching of excitonic
transitions and saturation of exciton population in NPL using two
color pump probe technique. This paper is organized as usual: In
Sec.~\ref{sec:synthesis} we describe the procedure of colloidal
synthesis of NPLs and their structure properties. Then in
Sec.~\ref{sec:exp} we present experimental setup and experimental
results, which are theoretically analyzed in Sec.~\ref{sec:theory}.
Comparison of theoretical predictions and experimental results
allowed us to estimate parameters of exciton dynamics including
average exciton lifetime and radiative recombination rate. Our main
experimental and theoretical findings are summarized in concluding
Sec.~\ref{sec:conclusion}.

\section{Synthesis and structural characterization of nanoplatelets.}
\label{sec:synthesis}

CdSe NPLs having five monolayer ($5$ML) thickness were synthesized
by the modified method adapted from Ref.~\onlinecite{Ithurria}. In a
typical synthesizes $0.5$ mmol cadmium acetate, $0.2$ mmol of oleic
acid (OA) and $10$ mL of octadecene (ODE) were introduced into a
reaction flask. The mixture was degassed under magnetic stirring and
argon flow at $180^{o}$C during $30$ minutes. Then, the temperature
was raised to $210^{o}$C and $0.2$ mmol trioctylphosphine (TOP) Se
solution was quickly injected into the mixture. The growth time of
NPLs was $40$ minutes at $210^{o}$C. The mixture was then cooled
down to room temperature and $1$ mL of OA was injected.
As-synthesized CdSe NPLs were precipitated by adding an equal volume
of acetone and centrifugation at $7000$ rpm for $5$ min and washed
two times with acetone. Finally, NPLs precipitates were re-dispersed
in $6$ mL of hexane.

The synthesis of CdSe/CdS heterostructures was carried out by the
method of low-temperature layer-by-layer deposition of shell
material according to Ref.~\onlinecite{Ithurria_1}. Briefly, $1$ mL
of a solution CdSe NPLs in hexane and $1$ mL a freshly prepared
$0.1$ M solution of Na$_2$S in N-methylformamide (NMF) were mixed
and shaken for $1$ hour. At the same time, a transfer of
nanoparticles from the nonpolar hexane phase to the polar NMF phase
was observed, which means the exchange of oleic ligands at the
surface of NPLs by a monolayer of $S^{2-}$. The polar phase was
rinsed several times with hexane and then the NPLs were precipitated
by adding a mixture of acetonitrile and toluene ($1:1$ by volume)
followed by centrifugation at $7000$ rpm and redispersed in NMF. The
precipitation-redispersion cycles were repeated two times to purify
the residues from unreacted sulfur-ions. Then $1$ mL of a solution
$0.3$ M of Cd(OAc)$_2$*2H$_2$O in NMF was added to the precipitate
and left for $40$ minutes to grow a CdS monolayer. The procedure
described above corresponds to the synthesis of a single CdS
monolayer. To obtain 5CdSe/2CdS heterostructures, this process was
repeated two times. The synthesized NPLs were investigated
experimentally in a liquid solution of methylformamide.

\begin{figure}
\includegraphics[width=\linewidth]{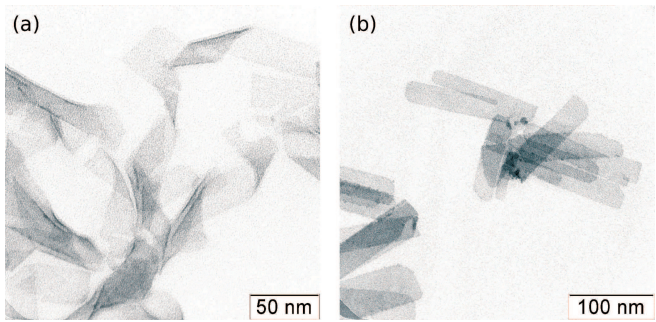}
\caption{Low-resolution transmission electron microscopy (TEM)
overview images of as-synthesized CdSe NPLs (a) and CdSe/CdS NPLs
with $2$ ML of CdS on each of basal planes~(b).} \label{fig:TEM}
\end{figure}

Low-resolution transmission electron microscopy (TEM) image of
as-grown CdSe NPLs [Fig.~\ref{fig:TEM}(a)] demonstrates
two-dimensional morphology. Rectangle platelets with lateral sizes
about $30\times 100$ nm rolled up into nanscrolls.
Fig.~\ref{fig:TEM}(b) presents typical image of CdSe NPLs after
covering with $2$ ML of CdS. We found that initially rolled NPLs
unfolded into flat nanostructures. As a result well-defined
rectangle CdSe/CdS platelets with the same lateral sizes as CdSe
NPLs were formed. The synthesized NPLs had zinc-blend crystal
structure~\cite{PhysRevB.89.035307}.

\section{Experiment}
\label{sec:exp}


\begin{figure}
\includegraphics[width=\linewidth]{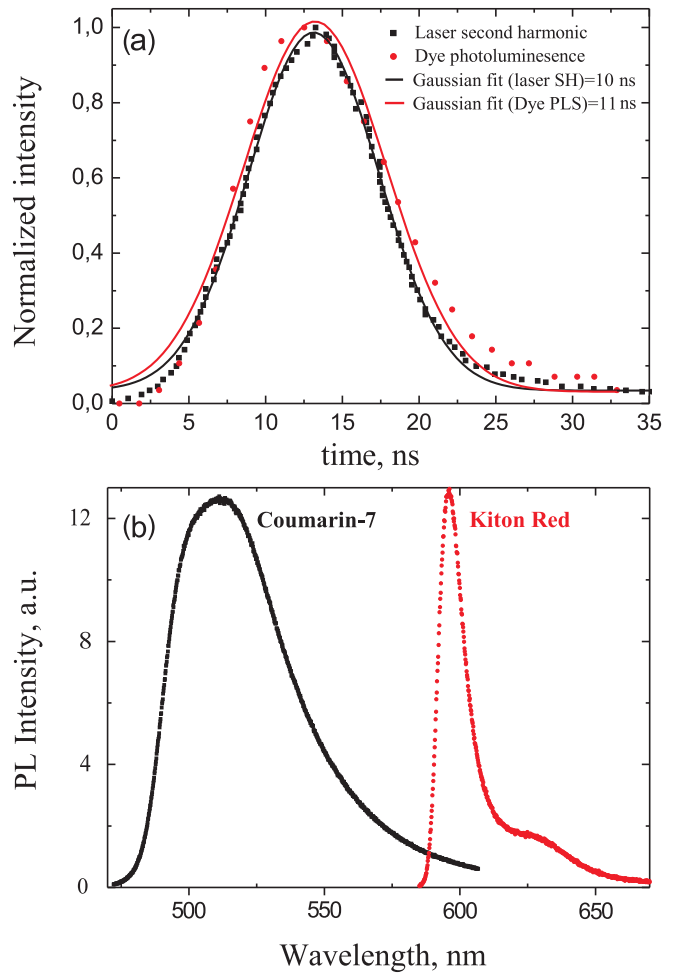}%
\caption{(a) The temporal profiles of the pump (red dots) and the
probe (black dots) pulses. The curves show the corresponding
Gaussian fits. (b) Spectra of the probe pulse, being the
photoluminescence spectra of Coumarin-$7$ or Kiton Red dyes.}
\label{figure_3}
\end{figure}

\begin{figure}
\includegraphics[width=\linewidth]{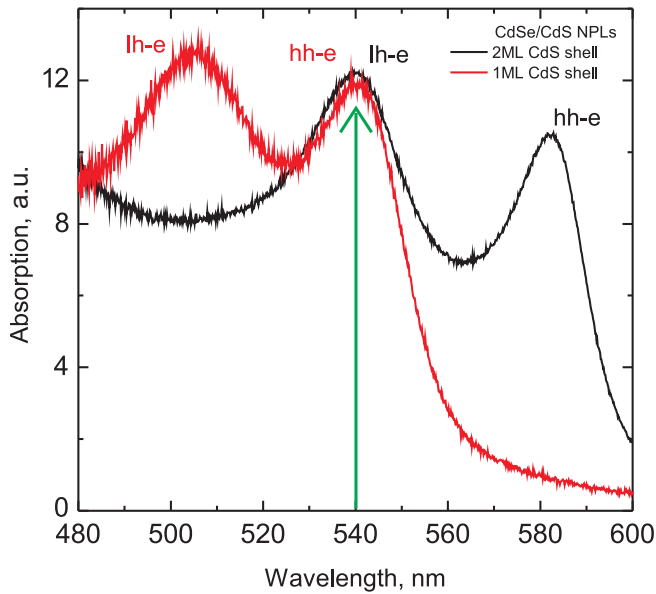}%
\caption{Absorption spectra of colloidal solution of 5CdSe/CdS and
5CdSe/2CdS NPLs. Green arrow shows the pump laser wavelength.}
\label{figure_4}
\end{figure}

\begin{figure}
\includegraphics[width=\linewidth]{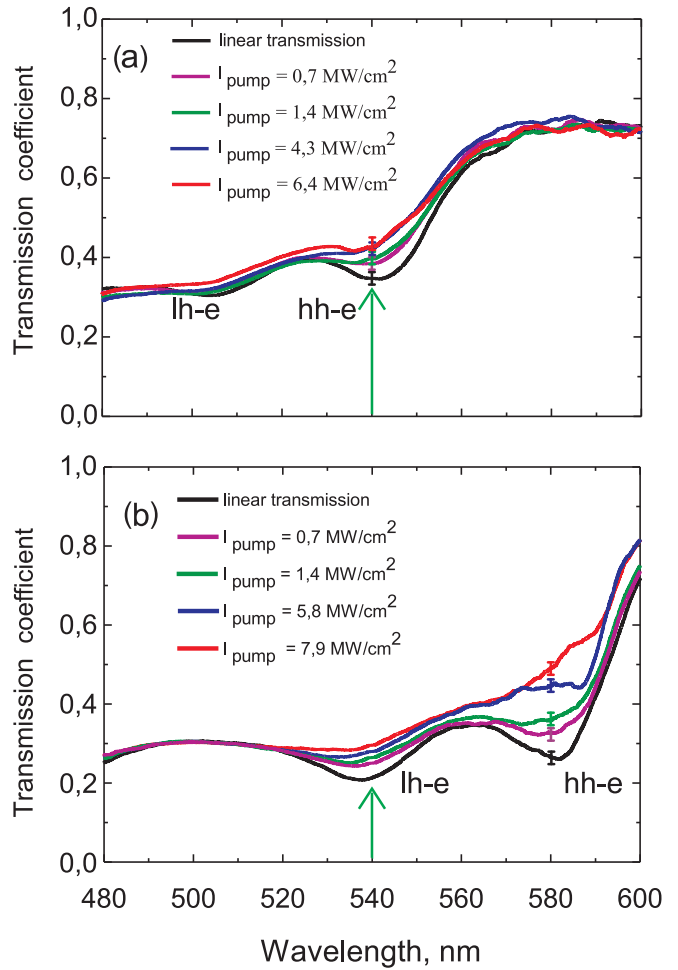}%
\caption{Transmission spectra of (a) 5CdSe/CdS and (b) 5CdSe/2CdS
NPLs. The green arrow corresponds to the pump wavelength. The
vertical bars demonstrate the characteristic statistical error.}
\label{figure_5}
\end{figure}

\begin{figure}
\includegraphics[width=\linewidth]{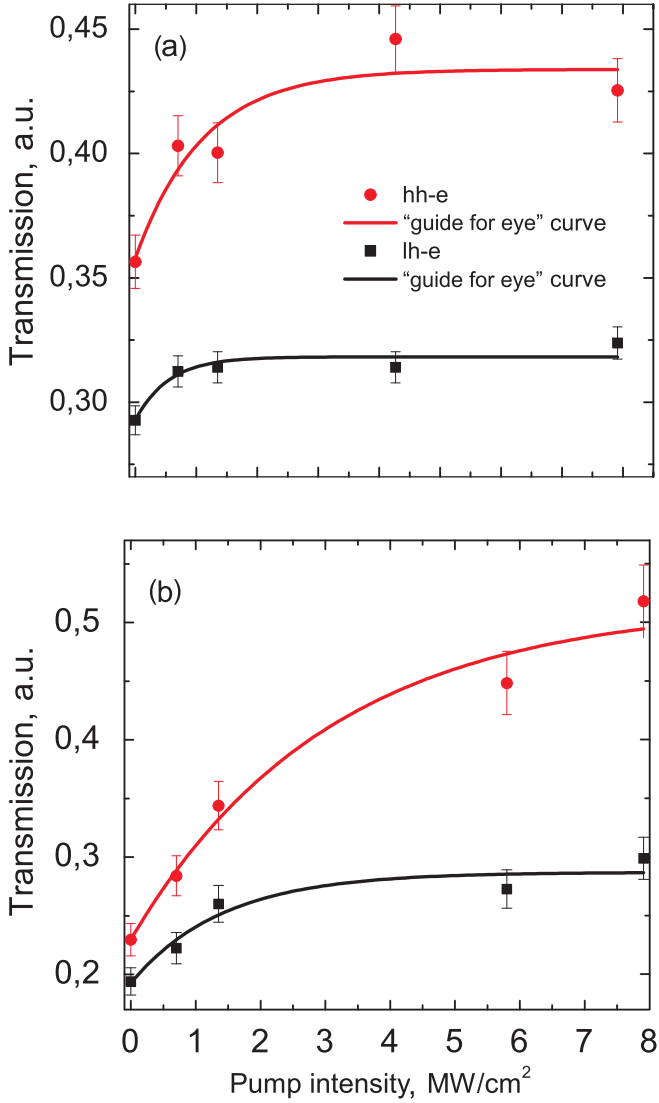}%
\caption{Transmission as a function of pump intensity for the (a)
5CdSe/CdS nanoplatelets and (b) 5CdSe/2CdS nanoplatelets.
Experimental dots correspond to the transmission spectra amplitudes
maxima obtained under resonant excitation (see Fig.\ref{figure_4}).}
\label{figure_6}
\end{figure}

\begin{figure}
\includegraphics[width=\linewidth]{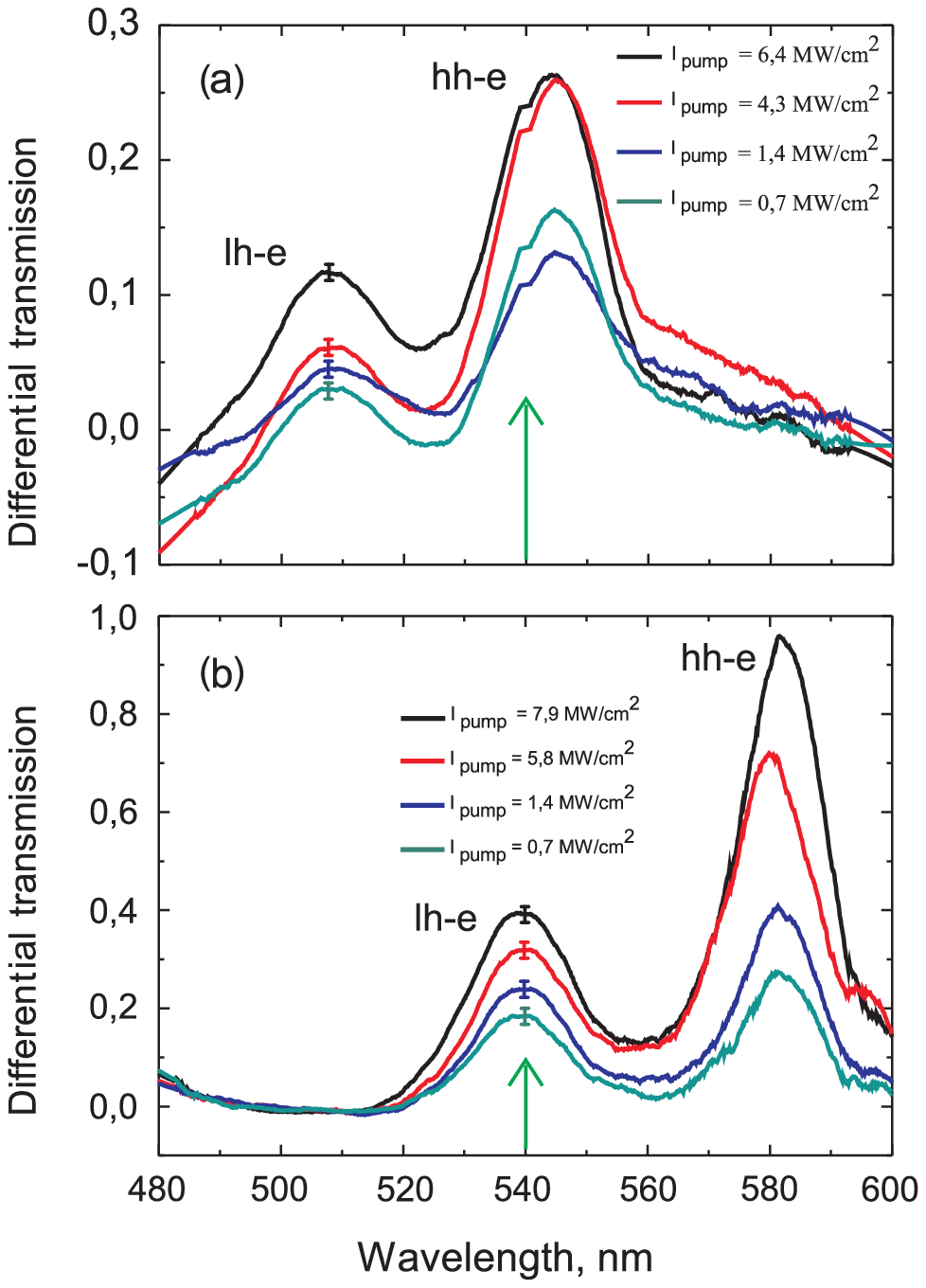}%
\caption{The differential transmission spectra calculated after
Eq.~\eqref{eq:DT} for the different pump powers. Panel (a)
corresponds to 5CdSe/CdS NPLs and panel (b) to 5CdSe/2CdS NPLs.
Laser excitation parameters correspond to the one, shown in
Fig.~\ref{figure_5}. The vertical bars demonstrate the
characteristic statistical error. } \label{figure_7}
\end{figure}

We study nonlinear optical properties of NPLs at room temperature
using pump-probe technique. Pumping was realized by the mode-locked
Nd$^{3+}$:YAlO$_3$ laser second harmonic ($\lambda=540$~nm, pulse
duration about $10$~ns). As a probe we used a broadband
photoluminescence radiation of the Coumarin-$7$ and Kiton Red dyes
excited by the third harmonic ($\lambda=360$~nm) of the pumping
laser~\cite{Dneprovskii}. The probe pulse duration was 11~ns, and we
adjusted it to overlap with the pump pulse, as shown in
Fig.~\ref{figure_3}(a). Because of relatively long pulse duration we
probe steady state of the system. The broad spectrum of the pump
light is shown in Fig.~\ref{figure_3}(b), it covers wavelengths from
$480$ to $600$~nm by Coumarin-$7$ photoluminescence and from $590$
to $660$~nm by Kiton Red photoluminescence.

The transmission and absorption of the probe light was measured with
spectral resolution using SpectraPro $2300i$ spectrometer with PIXIS
$256$ CCD-camera. The absorption spectra of 5CdSe/CdS and 5CdSe/2CdS
NPLs ($5$ core CdSe monolayers with $1$ or $2$ CdS shell monolayers,
respectively) in $1$~mm cell with methylformamide in the absence of
pumping are shown in Fig.~\ref{figure_4}. The two pronounced peaks
correspond to the heavy- and light-hole excitons~\cite{Kunneman}.
The energies of these transitions are determined by the size
quantization in direction perpendicular to NPLs. One can see, that
the two peaks shift in 5CdSe/2CdS NPLs with respect to 5CdSe/CdS
NPLs because of the deeper penetration of the exciton wavefunction
in CdS shell.

The wavelength of the pump laser is shown by the green arrow in
Fig.~\ref{figure_4}. One can see that it corresponds to heavy hole
(hh-e) and light-hole (lh-e) transitions in 5CdSe/CdS and 5CdSe/2CdS
NPLs, respectively. As a result the pump light resonantly excites
the corresponding excitonic states in the two studied samples.

The transmission spectra of the probe light for different pump
powers are shown in Fig.~\ref{figure_5}. Here one can see two dips
corresponding again to hh-e and lh-e transitions. Increase of the
pump power leads to the eventual disappearance of the dips.
Fig.~\ref{figure_6} demonstrates transmission as a function of pump
intensity for both samples at the hh-e and lh-e exciton transitions
wavelength.

We introduce the differential transmission at a given wavelength
$\lambda$ as
\begin{equation}
DT(\lambda)=\frac{T_{I}(\lambda)-T_{0}(\lambda)}{T_{0}(\lambda)},
\label{eq:DT}
\end{equation}
where $T_{I}(\lambda)$ is the transmission of the solution of
colloidal NPLs under pumping with intensity $I$ at the wavelength
$\lambda$. The differential transmission spectra are shown in
Fig.~\ref{figure_7}. They directly reflect nonlinear optical
properties of colloidal NPLs. In particular we note that under
strong excitation of heavy hole excitons [panel (a)] the
differential transmission relevant to the same transition saturates.
This suggests to develop a quantitative description of the observed
phenomena.

\section{Theory and discussion}
\label{sec:theory}

\begin{figure}[h!]
\centering
\includegraphics[width=0.6\linewidth]{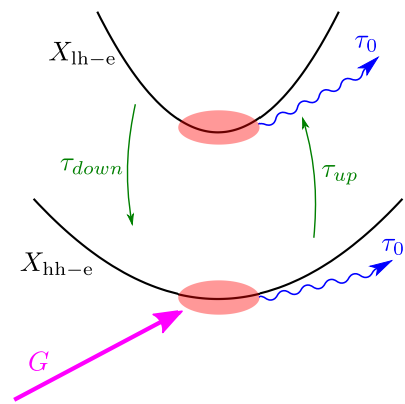}
\caption{Energy diagram of heavy hole ($X_{hh-e}$) and light hole
($X_{lh-e}$) excitons for resonant excitation of heavy-hole exciton.
The red ovals show the part of exciton spectrum, where radiative
decay takes place.} \label{fig:levels}
\end{figure}

The optical transitions related to hh-e and lh-e are well resolved
even at room temperature (see Fig.~\ref{figure_4}). This evidences
large exciton binding energy and strong oscillator strength due to
quantum confinement in perpendicular to NPL direction. In the
previous section we presented results of nonlinear absorption of
light under resonant excitation of heavy-hole and light-hole
excitons. Interestingly in both cases the differential transmission
is nonzero at both resonances. This means that after resonant
excitation not only light hole excitons can scatter to the heavy
hole excitonic states with lower energies, but also the inverse
process takes place. As a result the system under study can be
qualitatively described by the scheme shown in
Fig.~\ref{fig:levels}. In our experiment we did not observe any
traces of trapping of charges at defect
sites~\cite{Kunneman,Selyukov}, so we do not take this effect into
account in this model.

The differential transmission can be related to three
mechanisms~\cite{doi:10.1063/1.371244} (i) phase space filling, (ii)
screening of Coulomb interaction or (iii) broadening of the
excitonic line. All three effects appear because of exciton-exciton
interaction. For the following analysis we assume, that (i)
differential transmission is proportional to the number of excitons
in the corresponding state, and (ii) the times of exciton scattering
between heavy-hole and light-hole states [$\tau_{down}$ and
$\tau_{up}$, see Fig.~\ref{fig:levels}] are the shortest timescales
in the system. These assumptions are made to obtain the simplest
theoretical description of the system, which can yield estimations
for the parameters of exciton dynamics. Moreover we note, that
exciton fine structure~\cite{Rodina} can be neglected at room
temperature~\cite{Kunneman}.

The ratio of areas of differential transmission peaks corresponding
to heavy- and light-hole excitons is
\begin{equation}
  \frac{A_{hh}}{A_{lh}}=\frac{N_{hh}}{N_{lh}},
\end{equation}
where $N_{hh}$ and $N_{lh}$ are the concentrations of heavy- and
light-hole excitons in NPLs, respectively. The balance of
transitions between excitonic states reads
\begin{equation}
  \frac{N_{hh}}{\tau_{up}}=\frac{N_{lh}}{\tau_{down}}.
  \label{eq:N_rat}
\end{equation}
Accordingly one can find the ratio between the times of exciton
scattering to upper and lower energies as
\begin{equation}
\frac{\tau_{up}}{\tau_{down}}=\frac{A_{hh}}{A_{lh}}.
\end{equation}
This ratio is about $1.5$ for 5CdSe/CdS NPLs and about $2.5$ for
5CdSe/2CdS NPLs.

We note that $\tau_{up}/\tau_{down}$ is much smaller, than the
Boltzmann exponent for the corresponding splitting between
heavy-hole and light-hole excitonic states ($\approx 160$~meV).
Therefore we conclude, that exciton scattering is induced by
exciton-exciton interaction.

Under resonant excitation of heavy-hole excitons we observe
saturation of differential transition at corresponding wavelengths,
see Fig.~\ref{figure_7}(a). This effect is absent for resonant
excitation of light-hole excitons, which is in agreement with the
faster exciton scattering to lower than to higher energies,
$\tau_{down}<\tau_{up}$. Let us study the saturation effect in more
detail.

The saturation of differential transmission is related with the
saturation of heavy-hole excitons concentration
at~\cite{schmittrink85}
\begin{equation}
  N_s=\frac{7}{8\pi a_B^2},
\end{equation}
where $a_B$ is the 3D Bohr radius, which is two times bigger, than
the excitonic Bohr radius in narrow NPLs. The exciton
generation rate $G$ is given by

\begin{equation}
  G=\frac{P_S}{\hbar\omega_0}\frac{2\Gamma_0}{\Gamma},
  \label{eq:G}
\end{equation}
where $\omega_0$ is the pump carrier frequency and
$2\Gamma_0/\Gamma$ is the absorbance of the nanoplatelet with
$\Gamma_0$ and $\Gamma$ being, respectively, radiative and
nonradiative damping rates~\cite{ivchenko05a}.

The radiative exciton decay can take place only in the so-called
radiative cone --- an area in the momentum space with small enough
momenta~\cite{microcavities,Biadala}. This area is schematically
shown by red ovals in Fig.~\ref{fig:levels}. The exciton radiative
recombination time $\tau_0$ is given by~\cite{Yugova}
  \begin{equation}
    \frac{1}{\tau_0}=2\Gamma_0=\frac{4q^3}{3\varepsilon_b\hbar}|\mathcal D|^2,
    \label{eq:gamma0}
  \end{equation}
where $\varepsilon_b$ is the dielectric constant of the solution~\cite{Platonov}, $q=\sqrt{\varepsilon_b}\omega_0/c$ is the light wavevector with $c$ being the speed of light in vacuum, and
\begin{equation}
  |\mathcal D|^2=\frac{4e^2E_pA}{\pi\omega_0^2m_0a_B^2}.
  \label{eq:D}
\end{equation}
Here $e$ is the electron charge, $E_P$ is the energy parameter
measuring the Kane interband momentum matrix element, $A$ is the
area of NPL, $m_0$ is the free electron mass, and it is assumed that
$Aq^2\ll1$.

Finally the average exciton lifetime $\tau$ can be estimated from
the balance of exciton generation and recombination rates as
\begin{equation}
  G=\frac{N_{hh}+N_{lh}}{\tau}.
\end{equation}
Under saturation conditions one finds
\begin{equation}
  \tau=\frac{N_s}{G}\left(1+\frac{\tau_{down}}{\tau_{up}}\right),
  \label{eq:tau}
\end{equation}
where we have taken into account Eq.~\eqref{eq:N_rat}. For
estimations we use the following parameters: $\hbar\omega_0=2.3$~eV,
$\varepsilon_b=2$, $A=3000$~nm$^2$,
$a_B=5.6$~nm~\cite{doi:10.1021/nn3014855} and
$E_P=23$~eV~\cite{PhysRevB.78.035207}. We obtain
$\hbar\Gamma_0=87~\mu$eV, which corresponds to radiative lifetime of
an exciton at rest (with zero in plane momentum) $\tau_0=3.8$~ps.
The short radiative lifetime is related to relatively large area of
NPLs, see Eq.~\eqref{eq:D}. From Fig.~\ref{figure_6} one can also
estimate $P_S=4$~MW/cm$^2$ and the exciton linewidth
$\hbar\Gamma=40$~meV, which from Eqs.~\eqref{eq:G}
and~\eqref{eq:tau} gives the average exciton lifetime $\tau\approx
24$~ps in agreement with Ref.~\cite{Biadala}. We recall, that the
significant difference between $\tau_0$ and $\tau$ is related to the
fact that the radiative decay of excitons is possible only inside
the narrow radiative cone.

\section{Conclusion}
\label{sec:conclusion}

We studied experimentally nonlinear absorption and transmission of
light in ensembles of CdSe/CdS NPLs. We used the two-color
pump-probe technique with long pulse duration, which allowed us to
study quasi steady state. Resonant excitation of heavy- and
light-hole excitonic states revealed surprisingly small ratio of
exciton scattering times to higher and lower energies. This is
related to the nonlinear exciton scattering induced by
exciton-exciton interaction.

We also observed saturation of differential transmission at
heavy-hole exciton resonance. Theoretical analysis of this
phenomenon allowed us to obtain average exciton lifetime and
radiative recombination rate at room temperature. We believe, that
investigation of nonlinear optical properties of colloidal NPLs
performed in this paper will give a push to development of
optoelectronic devices operating at room temperature on the basis of
colloidal NPLs.

\section{Acknowledgements}

We thank E. L. Ivchenko and A. A. Golovatenko for fruitful
discussions and acknowledge the support by the RFBR grants
$16-29-11694$ and $18-02-00719$. D.S.S. was partially supported by
the Russian Foundation for Basic Research (Grant No. 17-02-0383), RF
President Grant No. SP-643.2015.5, and the Basis Foundation.

\end{document}